# Correcting Low-Signal Sensitivity in the Deliberative Reason Index

Francesco Veri, Centre for Democracy Studies, University of Zurich

*April 2026*

*Working Paper*

**Abstract**

The Deliberative Reason Index (DRI) is increasingly used to assess the coherence between considerations and preferences in deliberative settings, including applications to LLM-generated data. Under low-signal conditions, however, the standard DRI can produce inflated scores by treating near-zero correlations as evidence of consistency. Monte Carlo simulations across common study designs show that this bias increases with group size and yields positive values even under random response. A modified DRI is introduced that applies a continuous penalty to low-signal correlation pairs. The modification preserves the original scale and reduces exactly to the standard DRI when substantive signal is present. A threshold sensitivity analysis identifies $\tau = 0.2$ as the optimal parameter. An empirical check with archival deliberative data shows that substantive inferences remain unchanged. The modification improves the reliability and comparability of the DRI in low-signal settings.

## 1. Introduction

Deliberative reasoning is central to the normative aspirations of mini-publics (Niemeyer et al., 2024; Veri and Niemeyer, 2025). The Deliberative Reason Index (DRI), developed by Niemeyer and Veri (2022), operationalises this concept by measuring the intersubjective consistency between participants' ratings of considerations—how much weight they assign to a set of considerations on a Likert scale—and their ranked preferences over policy alternatives. High DRI values indicate that preferences are coherently grounded in stated considerations, whereas low values signal a disconnect. The index has been applied in assessments of deliberative quality in both experimental and real-world mini-public settings.

As the DRI is increasingly used to evaluate LLM-generated responses (e.g., Flechtner, 2026), it becomes important to assess its behaviour under low-signal conditions. This note builds on a formulation already applied in recent work (e.g., Umbelino and Veri, 2025) and implemented in the R package *deliberr* (10.32614/CRAN.package.deliberr), and provides a formal clarification and refinement of an underexplored aspect of the standard DRI calculation, enhancing its robustness in such settings.

The issue concerns the behaviour of the DRI when both the consideration correlation $r$ and the preference correlation $q$ for a given pair are close to zero. In such cases, the orthogonal distance $d = |r - q|/\sqrt{2}$ is mechanically small, even though both correlations are negligible. Because the standard formulation aggregates low distances as evidence of intersubjective consistency, these pairs contribute positively to the index despite containing little informational content. As a result, the aggregate DRI may be inflated.

While this behaviour is typically not empirically salient in conventional deliberative datasets—where some degree of structured reasoning is almost always present—it becomes relevant under conditions where responses are weakly structured or effectively random. This includes settings characterised by low engagement, survey fatigue, or, more importantly, LLM-generated data, where response patterns may not reflect a stable underlying structure. Monte Carlo simulations confirm that, under fully random responses, the standard DRI produces mean values of approximately 0.30-0.6, depending on group size—levels that could be misinterpreted as evidence of deliberative quality.

This note therefore formalises and justifies a correction to the standard DRI calculation that downweights low-signal correlation pairs. The proposed modification preserves the original scale and interpretation of the index while extending its robustness to settings in which weak or random response structures are more likely to occur, including the analysis of LLM-generated data.

The proposed modification introduces a continuous scalar penalty that reduces the contribution of low-signal pairs in proportion to their distance from the origin of the $(r, q)$ correlation space. The penalty is parameterised by a threshold $\tau$: pairs with $\max(|r|, |q|) < \tau$ receive reduced weight, while pairs above $\tau$ receive full weight. A systematic sensitivity analysis across four candidate threshold values ($\tau \in \{0.1, 0.2, 0.3, 0.4\}$) demonstrates that $\tau = 0.2$ is the optimal choice on empirical and substantive grounds: it is the highest value that keeps the noise floor close to zero rather than pushing it into negative territory, and it aligns with the conventional boundary of negligible correlation in groups of the size typically assembled for deliberative research.

The paper is structured as follows. Section 2 restates the standard DRI in its lambda-based form. Section 3 derives the modified formula. Section 4 presents Monte Carlo validation of the standard and modified DRI across study designs (Component A). Section 5 reports the threshold sensitivity analysis (Component B). Section 6 presents an empirical check, Section 7 discusses limitations. Section 8 concludes.

## 2. The Standard DRI: Formulation and Scaling Correction

### 2.1. Measurement Structure

The DRI is applied to data from a deliberative group of $n$ respondents who have rated $C$ consideration questions (using Likert scales or quasi-ranked responses based on Q methodology; Stephenson 1959) and ranked $P$ preference alternatives. For each consideration $c \in \{1, ..., C\}$ and each preference $p \in \{1, ..., P\}$, an across-respondent correlation $r_{cp}$ is computed between the consideration rating vector and the preference rank vector. This yields a $C \times P$ matrix of correlations.

In two-wave or subgroup applications of the DRI, correlations are computed separately for each wave or subgroup, such that each cell contains a pair $(r_{cp}, q_{cp})$. The DRI then summarises the agreement between these two relational structures.

### 2.2. Orthogonal Distance

For each $(r, q)$ pair, the orthogonal distance from the main diagonal—representing perfect agreement—is:

$$d = \frac{|r - q|}{\sqrt{2}}$$

This corresponds to the perpendicular distance from the point $(r, q)$ to the line $r = q$ in the unit square $[-1,1]^2$, consistent with the geometric formulation of intersubjective consistency

### 2.3. DRI Scale Transformation

Distances are first averaged across all $C \times P$ pairs to obtain $\bar{d}$, and then transformed to a bounded scale using the lower-bound constant $\lambda$, which represents the theoretical maximum average distance for a given group:

$$DRI = \frac{-2\bar{d} + \lambda}{\lambda}$$

This formulation follows directly from the individual-level transformation proposed by Niemeyer and Veri (2022), where DRI is obtained by scaling average orthogonal distance relative to its theoretical maximum

## 3. The Modified DRI: Scalar Penalty Mechanism

### 3.1. The Inflation Problem

When both $r$ and $q$ for a given pair are near zero, their distance $d = |r - q|/\sqrt{2}$ is small regardless of whether these near-zero values reflect genuine agreement or mutual noise. Because the DRI aggregates low distances as evidence of consistency, such pairs contribute positively to the index even though both correlations are negligible. As a result, the aggregate DRI may be inflated.

The geometric intuition is straightforward. In the $(r, q)$ unit square, the distance $d$ measures proximity to the diagonal $r = q$. The problematic region is the neighbourhood of the origin $(0, 0)$, where both correlations are negligible and $d$ is therefore small—not because the group is reasoning coherently, but because both correlations are close to zero.

### 3.2. Scalar Penalty

The penalty assigns each $(r, q)$ pair a weight in $[0, 1]$ based on the maximum absolute value of the two correlations:

$$\text{penalty} = \begin{cases} \dfrac{\max(|r|, |q|)}{\tau}, & \text{if } \max(|r|, |q|) \leq \tau \\ 1, & \text{otherwise} \end{cases}$$

Pairs for which neither correlation exceeds $\tau$ in absolute value receive a weight proportional to their signal strength. The weight increases linearly from zero at the origin to one at the threshold boundary. The function is continuous at $\tau$, avoiding any discontinuity in weighting, and ensures that pairs at or above the threshold receive full weight. Pairs above the threshold—where at least one correlation is substantively non-negligible—are therefore unaffected.

### 3.3. Adjusted Consistency and Final Formula

Each pair's contribution is adjusted by applying the penalty to the distance:

$$d^* = d \times \text{penalty}$$

The adjusted average distance is then:

$$\bar{d}^* = \text{mean}(d^*)$$

The DRI is computed using the standard transformation:

$$DRI = \frac{-2\bar{d}^* + \lambda}{\lambda}$$

When all pairs satisfy $\max(|r|, |q|) > \tau$, the penalty equals 1 for all observations and the modified formulation reduces exactly to the standard DRI. The adjustment is therefore dormant in the presence of substantive signal and becomes active only when low-signal pairs are present.

## 4. Component A: Main Validation

### 4.1. Simulation Design

Component A validates the standard and modified DRI across a range of study designs typical in deliberative research, holding group size constant at n=30. Synthetic data were generated by crossing three numbers of considerations (15, 30, 50), two numbers of preference alternatives (4, 10), and two Likert scale formats (1–5, 1–7), yielding 12 design conditions. Each condition was combined with five noise levels (0, 0.25, 0.50, 0.75, 1.0) and replicated 1000 times, producing 60,000 simulated datasets.

A second simulation was conducted with group size held constant at $n = 100$, using the same design structure and replication scheme. This yields an additional 60,000 simulated datasets, for a combined total of 120,000 simulations across both group-size conditions.

The noise parameter controls the proportion of responses drawn from a uniform random distribution rather than from a shared latent structure linking considerations and preferences. At noise = 0, all responses are generated from the latent structure, producing coherent correlation patterns; at noise = 1, all responses are uniformly random. Intermediate values represent mixtures of structured and random responses.

Data were generated using a split-half design: each group of n=30 was randomly divided into two halves, with consideration–preference correlations r computed from the first 15 respondents and q from the second 15, mirroring the subgroup structure in which DRI is typically applied.

### 4.2. Results

Figure 1 presents the mean DRI by formula and noise level, averaged across all design conditions. Three findings stand out.

First, at noise = 0 and noise = 0.25, the two formulas return identical values within each group-size condition, confirming that the penalty mechanism is fully dormant when genuine signal is present. The bias row shows 0.000 at both levels, confirming that the penalty mechanism is fully dormant when genuine deliberative signal is present. This is a necessary property: the modification leaves high-quality deliberative scores unchanged.

Second, the formulas diverge sharply beyond noise = 0.5. The standard DRI continues to decline but stabilises at approximately 0.394 for n= 30 and 0.677 for n=100 under pure randomness—a value that does not meaningfully distinguish between structured reasoning and random responding. By contrast, the modified DRI declines more steeply, reaching 0.132 at noise = 1 for n=30 and -0.094 for n=100.

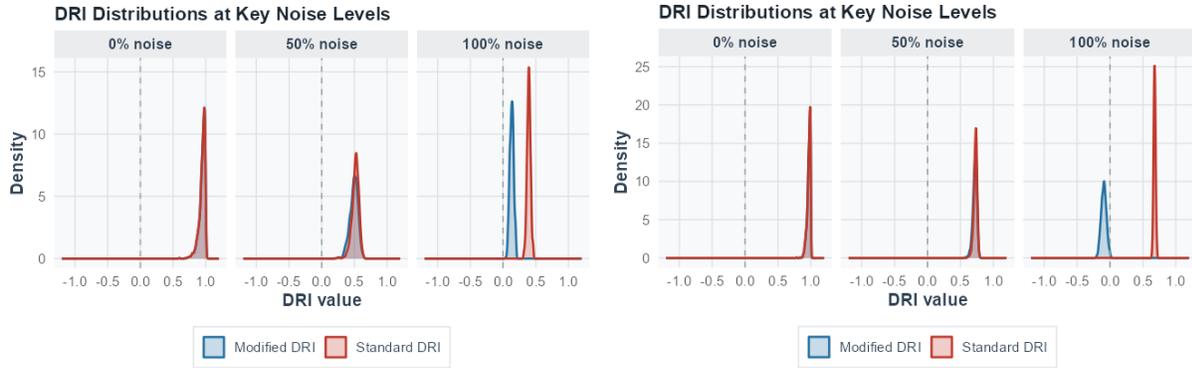

**Figure 1** : *DRI formula under randomly generated data (0% = not random / 100% = fully random data) – n=30 left plot, n=100 right plot.*

## 4.3. Design Invariance

A secondary finding concerns the robustness of the modified DRI noise floor across study design parameters. At noise = 1, the mean modified DRI is approximately 0.13 for $n = 30$ and approximately −0.10 for $n = 100$, regardless of whether the study includes 15 or 50 consideration questions, 4 or 10 preference alternatives, or a 5- or 7-point Likert scale.

Taken together, these values define a narrow range centred around zero, indicating that the lower bound of the modified DRI remains stable across both instrument design and group size. This stability reduces sensitivity to survey design choices and participant count.

As a result, researchers can interpret values close to zero—whether slightly positive or slightly negative—as indicative of the absence of structured signal, without requiring instrument-specific calibration.

## 5. Component B: Threshold Sensitivity Analysis

### 5.1. Design

The choice of threshold $\tau$ requires both empirical grounding and substantive justification. Component B systematically compares four candidate values $\tau \in \{0.1, 0.2, 0.3, 0.4\}$ across 150 scenarios (4 thresholds × 5 noise levels × 6 design conditions), with 300 simulations per scenario. Four evaluation criteria are considered:

(1) *Discrimination*: $DRI_{noise=0} - DRI_{noise=1}$, where higher values indicate greater separation between fully structured and fully random responses.

(2) *Noise floor*: the mean DRI at noise = 1, representing the value assigned to fully random responses. The ideal is close to zero—low enough to indicate absence of structured signal, but not so negative as to invert the interpretation of the scale.

(4) *Fidelity*: | DRI$_{modified}$ − DRI$_{standard}$ | at noise = 0, where lower values indicate closer alignment with the standard DRI under conditions of genuine signal.

### 5.2. Results

Table 1 presents the threshold criteria. Two patterns are immediately apparent.

First, discrimination increases monotonically from 0.545 at $\tau = 0.1$ to 1.825 at $\tau = 0.4$. This pattern is expected: larger thresholds penalise a greater share of low-signal pairs, lowering the noise floor and increasing the separation between structured and random responses. If discrimination were considered in isolation, higher thresholds would be preferred.

Second, at $\tau = 0.1$, the noise floor is 0.429—substantially positive and difficult to distinguish from a genuinely low-signal group. At $\tau = 0.2$, the floor around 0 (-0.096), and readily interpretable as the absence of structured signal. At $\tau = 0.3$, the floor becomes strongly negative (−0.552), such that fully random-response groups receive negative DRI scores.

This transition is critical for threshold selection. Once the noise floor becomes negative, the interpretation of the DRI scale is altered: the midpoint (0) no longer corresponds to a null signal baseline. Negative values would therefore require a revised interpretive framework not implied by the original index.

Table 1. Threshold Sensitivity Criteria (n = 100, averaged over design conditions)

| τ | Discrimination | Noise floor | Fidelity gap | Floor ≈0? | Monotone | Assessment |
|---|---|---|---|---|---|---|
| 0.1 | 0.545 | 0.429 | 0.000 | No | Yes | Under-penalises; floor still misleadingly positive |
| **0.2** | **1.065** | **-0.096** | **0.000** | **Yes ✓** | **Yes** | Recommended. Near-zero floor; preserves weak signal |
| 0.3 | 1.521 | −0.552 | 0.000 | No | Yes | Over-penalises; random groups score below zero |
| 0.4 | 1.825 | −0.855 | 0.000 | No | Yes | Strong over-penalty; penalises moderate signal |

### 5.3. The Case for τ = 0.2

The 0.2 level is recommended on two converging grounds.

*Empirical ground: interpretive integrity*. The noise floor at $\tau = 0.2$ with n=100 is -0.096, close to zero. This implies that a group of 100 respondents answering entirely at random would receive a modified DRI of approximately -0.096 —a value that can be interpreted as near-null signal. At $\tau = 0.3$, the floor becomes strongly negative (−0.552), such that random-response groups score below zero. This introduces an interpretive complication, as zero conventionally marks the boundary between absence and presence of positive association in correlation-based measures. Thresholds at $\tau \geq 0.3$ would therefore require rescaling or re-norming to preserve the zero-as-null interpretation.

*Conventional ground: Cohen's benchmark*. The threshold $\tau = 0.2$ lies between Cohen's (1988) conventional effect size benchmarks for correlations—0.10 (small), 0.30 (medium), and 0.50 (large). While Cohen does not define a threshold at 0.2, this value can be interpreted as a conservative boundary separating negligible from weak associations. The penalty function thus operationalises a standard methodological convention: correlations below this threshold are discounted, while those at or above it receives full weight. The penalty function thus operationalises a standard methodological convention: correlations below the negligible threshold are discounted, while those at or above it receives full weight. This provides a field-neutral justification that does not depend on deliberation-specific assumptions.

Taken together, these considerations support $\tau = 0.2$ as the threshold that best balances discrimination, interpretive integrity, and methodological convention. Researchers working with substantially larger samples or alternative theoretical assumptions may adjust $\tau$ accordingly, but should note that higher thresholds risk pushing the noise floor below zero and altering the interpretation of the DRI scale.

## 6. Empirical check

To complement the simulation analysis, the modified DRI is applied to a set of archival deliberative cases used in Veri (2025). This provides a first empirical check of how the adjustment behaves in observed data, where some degree of structured signal is present.

Across all cases, the modified DRI yields values that are very close to those produced by the standard formulation, with only small downward adjustments (Table 4). Pre–post differences and their statistical significance remain substantively unchanged. For example, in Winterthur, the standard DRI increases from 0.22 to 0.54 (Δ = 0.33***), while the modified DRI increases from 0.22 to 0.52 (Δ = 0.30***). Similar patterns are observed across all cases, indicating that the modification does not alter substantive inference when applied to genuine deliberative settings.

Table 2. Empirical check of the new formula

| case | N | Pre | Post | delta | pre | post | delta | Delta indexes pre | Delta indexes post |
|---|---|---|---|---|---|---|---|---|---|
| Bellinzona | 8 | 0.45 | 0.64 | **0.19*** | 0.45 | 0.64 | -0.00 | **-0.00** | 0 |
| Swiss Health Cost | 56 | 0.26 | 0.27 | **0.01^** | 0.17 | 0.20 | **0.03^** | -0.09 | -0.06 |
| Uster | 15 | 0.39 | 0.52 | **0.12*** | 0.34 | 0.50 | **0.15*** | -0.05 | -0.02 |
| Winterthur | 16 | 0.22 | 0.54 | **0.33\*\*\*** | 0.22 | 0.52 | **0.30\*\*\*** | -0.00 | -0.02 |
| Zukunftrat | 63 | 0.43 | 0.53 | **0.10\*\*\*** | 0.42 | 0.50 | **0.08\*\*\*** | -0.01 | -0.03 |

Taken together, the empirical evidence suggests that the modified DRI operates as a conservative adjustment. It leaves results effectively unchanged in contexts where structured reasoning is present, while providing protection against the inflation bias identified in low-signal conditions. This property is particularly important for applications in which response quality cannot be independently verified, such as large-scale deliberative processes or LLM-generated data

## 7. Discussion and Limitations

The simulation results demonstrate that the standard DRI exhibits a measurable and substantively relevant inflation bias under low-signal conditions. The modified DRI with $\tau = 0.2$ mitigates this bias while preserving the behaviour of the original formulation under genuine deliberative signal. One limitation of the analysis warrant explicit acknowledgment.

*Random response as the noise model.* The simulation evaluates formula behaviour under uniform random response, representing one end of a broader spectrum of non-deliberative patterns. Other response processes—such as systematic agreement irrespective of content, or midpoint or first last-option selection—can generate correlation structures that differ from uniform noise.

## 8. Conclusion

The standard DRI calculation exhibits a structural inflation bias under low-signal conditions: it assigns elevated scores when pairwise correlations are close to zero, a pattern that arises under random or disengaged response processes and becomes more pronounced as group size increases. The modified DRI with $\tau = 0.2$ returns approximately 0 for various simulations, correctly indicating the absence of structured signal.

Adoption of the modified DRI is therefore recommended, particularly in settings where response quality cannot be independently verified particularly in LLMs generated dataset. The modification reduces to the standard DRI when signal is present, requires no instrument-specific calibration, and resolves a previously undocumented group-size dependence that leads the standard formulation to produce increasingly inflated scores in larger deliberative assemblies.